\documentclass[pdflatex,sn-mathphys-num]{sn-jnl}

\usepackage{graphicx}%
\usepackage{multirow}%
\usepackage{amsmath,amssymb,amsfonts}%
\usepackage{amsthm}%
\usepackage{mathrsfs}%
\usepackage[title]{appendix}%
\usepackage{xcolor}%
\usepackage{textcomp}%
\usepackage{manyfoot}%
\usepackage{booktabs}%
\usepackage{algorithm}%
\usepackage{algorithmicx}%
\usepackage{algpseudocode}%
\usepackage{listings}%
\usepackage{rotating}%
\usepackage{caption}%

\theoremstyle{thmstyleone}%

\theoremstyle{thmstyletwo}%

\theoremstyle{thmstylethree}%

\begin{document}

\title[Article Title]{Greenhouse Gas (GHG) Emissions Poised to Rocket: Modeling the Environmental Impact of LEO Satellite Constellations}

\author[1]{\fnm{Rushil} \sur{Kukreja}}
\author[1]{\fnm{Edward J.} \sur{Oughton}}
\author[2]{\fnm{Richard} \sur{Linares}}

\affil[1]{\orgname{George Mason University}, \orgaddress{\street{4400 University Dr}, \city{Fairfax}, \state{VA}, \country{USA}}}
\affil[2]{\orgname{Massachusetts Institute of Technology}, \orgaddress{\street{77 Massachusetts Ave}, \city{Cambridge}, \state{MA}, \country{USA}}}

\abstract{The proliferation of satellite megaconstellations in low Earth orbit (LEO) represents a significant advancement in global broadband connectivity. However, we urgently need to understand the potential environmental impacts, particularly greenhouse gas (GHG) emissions associated with these constellations. This study addresses a critical gap in modeling current and future GHG emissions by developing a comprehensive open-source life cycle assessment (LCA) methodology, applied to 10 launch vehicles and 15 megaconstellations. Our analysis reveals that the production of launch vehicles and propellant combustion during launch events contribute most significantly to overall GHG emissions, accounting for 72.6\% of life cycle emissions. Among the rockets analyzed, reusable vehicles like Falcon-9 and Starship demonstrate 95.4\% lower production emissions compared to non-reusable alternatives, highlighting the environmental benefits of reusability in space technology. The findings underscore the importance of launch vehicle and satellite design choices to minimize potential environmental impacts. The Open-source Rocket and Constellation Lifecycle Emissions (ORACLE) repository is freely available and aims to facilitate further research in this field. This study provides a critical baseline for policymakers and industry stakeholders to develop strategies for reducing the carbon footprint of the space industry, especially satellite megaconstellations.}


\maketitle

\section{Introduction}\label{sec1}

Climate change poses an existential threat to the planet, with its impacts already evident across society and the economy \cite{huggel_existential_2022, vicedo-cabrera_burden_2021}. Economically, the projected cost of climate change is substantial, with one estimate placing a decline of 11\%–29\% in global income within the next 26 years, irrespective of future emission reductions \cite{kotz_economic_2024}. Extreme weather events exacerbated by climate change impose substantial global costs, on businesses, households and governments, with current estimates at \$143 billion annually, with 63\% of these losses attributed to human mortality \cite{newman_global_2023}. Additionally, climate-induced disruptions could reduce gross domestic product by as much as 10\% on average \cite{waidelich_climate_2024}. These statistics underscore the urgency of adopting ambitious mitigation and adaptation strategies to address climate impacts.

While the environmental impact of human activities on terrestrial ecosystems and the atmosphere has been extensively studied, the sustainability and techno-economics of space activities remains underexplored \cite{osoro_techno-economic_2021}. Satellite megaconstellations, such as SpaceX's Starlink and Amazon's Project Kuiper, hold the potential to revolutionize global communication by providing broadband internet to underserved regions, thereby narrowing the digital divide \cite{del_portillo_technical_2019, giannopapa_space_2022}, especially via implementation in developing countries \cite{wood_charting_2012}. Additionally, space technologies play a crucial role in climate monitoring, disaster response, and Earth observation, offering invaluable data for tracking deforestation, ice sheet melting, and greenhouse gas (GHG) emissions \cite{wentz_precise_2000, dinas_innovative_2015}. These benefits demonstrate that space is not inherently detrimental to the environment but rather a domain with both significant opportunities and challenges.

However, as the scale of these systems continues to expand, with proposals for constellations now exceed 100,000 satellites \cite{falle_one_2023}, concerns about their environmental impact and threats to space sustainability have grown. Satellite megaconstellations introduce significant challenges, including increased GHG emissions from frequent rocket launches, risks to atmospheric chemistry from propellant combustion, and the proliferation of space debris in low Earth orbit (LEO). The CO$_2$ emissions from the frequent rocket launches required to deploy and maintain these constellations are projected to grow significantly, amplifying their environmental impact and necessitating urgent scrutiny as launch frequencies rise \cite{dallas_environmental_2020}. These concerns call for a detailed evaluation of the life cycle emissions associated with satellite megaconstellations and the development of strategies to mitigate their environmental footprint \cite{osoro_sustainability_2024}.

This study distinguishes itself by conducting a comprehensive life cycle assessment (LCA) of the environmental impact of satellite megaconstellations. Leveraging satellite orbital licensing data, this research quantifies emissions from the production, transportation, and deployment of satellites for current and proposed constellations, considering a range of potential rocket launch vehicles. By aiming to quantify a variety of environmental metrics for these satellite systems, this study provides a detailed understanding of their impact across different lifecycle stages. Furthermore, the findings are presented as an open-source dataset, enabling transparency, reproducibility, and broader applications within the research community. Specifically, the research focuses on three primary questions:

\begin{enumerate}
    \item Rocket Emissions: What are the GHG emissions associated with each rocket used as a launch vehicle for satellite deployment?
    \item Constellation Emissions: What are the total emissions generated by each satellite megaconstellation, accounting for production, transportation, and launches?
    \item Per-User Emissions: What are the potential emissions attributable to each user of these satellite services under different adoption scenarios?
\end{enumerate}

By addressing these questions, this study seeks to provide a comprehensive understanding of the environmental implications of satellite megaconstellations. The analysis aims to inform policymakers, industry stakeholders, and researchers, highlighting opportunities for technological innovation and sustainable practices in the rapidly growing space industry.

The structure of the paper unfolds as follows: Section II presents a comprehensive literature review. In Section III, the methodology is outlined, detailing the approach designed to address the stated research questions. Section IV reports the results, which are subsequently analyzed and discussed in Section V. Finally, Section VI concludes the paper, summarizing key findings and their implications.

\section{Literature Review}\label{sec2}

There have been growing concerns around the use of near-Earth space \cite{rossi_future_2024, dambrosio_capacity_2022, sanchez-ortiz_outreach_2025}. The space industry has contributed to unprecedented growth in orbital activity, with the number of active satellites rising to approximately 7,400 by 2023, prompting increased scrutiny of environmental impacts \cite{jones_green_2023}. Lawrence et al. argue that the orbital space surrounding Earth should be regarded as an additional ecosystem, deserving the same level of care and protection as terrestrial ecosystems such as oceans and the atmosphere. They specifically emphasize the concerns for the sustainability of commercial, civic, and military activities in space. This perspective calls for a holistic approach to environmental stewardship that incorporates both terrestrial and orbital ecosystems \cite{lawrence_case_2022}. Recent research further underscores the urgency of managing orbital capacity, demonstrating that while very low Earth orbit (VLEO) could theoretically sustain hundreds of thousands to millions of satellites due to atmospheric drag, higher altitudes face increasing risks of debris accumulation, which could destabilize the orbital environment over centuries \cite{lavezzi_stable_2024}.

LEO satellites play a pivotal role in advancing global communication and navigation technologies, providing service levels unimaginable for remote locations a decade ago \cite{ahmmed_digital_2022, reid_broadband_2018}. Research continues to advance, with recent cellular generations, such as 6G, making non-terrestrial networks (NTNs) a central use case \cite{oughton_reviewing_2024, oughton_future_nodate, shahid_emerging_2024}. There are a range of benefits from LEO networks which increasingly provide high capacity, low latency wide-area coverage, enabling sustainable digital transformation for businesses and improvements in socio-economic opportunities for local communities \cite{liao_integration_2023}. Emerging LEO-HTS constellations, such as OneWeb and Starlink, aim to provide global broadband coverage, with Starlink planning up to 12,000 satellites in two orbital layers to achieve latencies as low as 10–15 ms, making LEO systems integral to the integration of space and terrestrial communication networks \cite{su_broadband_2019}. However, large satellite constellations also pose threats to the sustainability of the orbital environment \cite{bastida_virgili_risk_2016, dambrosio_carrying_2024}.

One of the most pressing environmental challenges stems from the rocket launches required to deploy and maintain large-scale satellite constellations \cite{ross_implications_2022}. Solid rocket propellants release CO$_2$ during combustion, comprising approximately 3.6\% of the total exhaust by mass, alongside other pollutants such as HCl and Al$_2$O$_3$, highlighting the contribution of these launches to atmospheric carbon levels \cite{noauthor_rocket_nodate}. Rocket launches in 2019 emitted 5,820 tonnes of CO$_2$, 6,380 tonnes of H$_2$O, 280 tonnes of black carbon, 220 tonnes of nitrogen oxides, 500 tonnes of reactive chlorine, and 910 tonnes of alumina into the stratosphere, highlighting the significant contributions of space activity to atmospheric pollution \cite{brown_worldwide_2024}. Consequently, satellite megaconstellation missions have emerged as a substantial and growing contributor to CO$_2$ emissions from space activities, with their share increasing from 26\% in 2020 to 33\% in 2022 \cite{barker_global_2024}.

While current launch rates are around 100 per year, mega-constellation deployment and maintenance will require dramatic increases. For scale, just the Starlink constellation represents about 3,100 tonnes of material that needs to be launched and replaced on 5-year cycles, resulting in approximately 2 tonnes of material reentering Earth's atmosphere daily.  Modeling suggests that an annual launch cadence of 1,000 hydrocarbon rocket launches could result in radiative forcing levels comparable to those from global subsonic aviation within a decade \cite{boley_satellite_2021}.

Additionally, rocket emissions of black carbon are up to 500 times more efficient at warming the atmosphere than surface and aviation sources, with sustained launches potentially contributing an additional 7.9 mW $m^{-2}$ to global radiative forcing, doubling the effect of contemporary rockets in just three years \cite{noauthor_impact_nodate}. These BC emissions are predicted to dominate atmospheric impacts by 2050 \cite{tsigaridis_composition_nodate}. The use of solid rocket boosters further introduces damaging particulates, such as alumina, directly into the upper atmosphere and stratosphere, contributing to atmospheric pollution and radiative forcing \cite{ross_radiative_2014}.

The emissions from solid rockets further exacerbate environmental concerns through ozone depletion in the stratosphere \cite{ross_observation_1997, donou-adonsou_space_2024}. Projections indicate that increased launch rates, necessary to deploy over 100,000 satellites by 2050, could lead to stratospheric ozone losses equivalent to 6\% of current annual global ozone depletion levels. This quantity is greater than the total emissions from now-banned ozone-depleting substances \cite{miraux_environmental_2022}. At these launch rates, hydrogen-fueled reusable rockets could increase global stratospheric water vapor by approximately 10\%, enhance polar mesospheric cloud fractions by 20\%, and reduce the globally averaged ozone column by 1.4–1.5 Dobson Units \cite{larson_global_2017}.

The scale of planned satellite megaconstellations also necessitates frequent replenishment launches to replace end-of-life satellites, further increasing the density of objects in LEO. Predictions suggest that an additional 20,000 rocket launches will occur in this decade alone, with between 60,000 and 100,000 satellites in LEO by mid-century \cite{noauthor_environmental_nodate}. Unlike traditional large GEO satellites, which are designed to last approximately 20 years, the mean lifespan of modern LEO satellites is significantly shorter, averaging only about 5 years. This shift in satellite lifespan introduces a step change in operational requirements, necessitating continuous replenishment to maintain the functionality of these constellations \cite{chou_towards_2022}. Even modest increases in launch traffic for constellation maintenance are projected to significantly amplify the risk of on-orbit collisions and debris generation \cite{somma_sensitivity_2019, zhang_leo_2022}. Additionally, satellite megaconstellations cause light pollution, negatively impacting the essential human right to dark skies and cultural sky traditions, affecting communities worldwide \cite{venkatesan_impact_2020}\cite{varela_perez_increasing_2023}. They also negatively impact telescope observations, potentially threatening the scientific viability of Earth-based telescopes. Simulations suggest that by 2030, 30\% of the images collected by some observatories could contain at least one satellite trail \cite{gwynne_observations_2020}.

The cumulative environmental effects of rocket launches—ranging from radiative forcing and ozone depletion to orbital crowding and alumina pollution—present significant challenges that must be addressed \cite{brown_envisioning_2024}. In addition, the production of a single Starlink V1 satellite is associated with a total environmental impact of 76 kilopoints, with resource use of minerals and metals contributing 58 kilopoints, emphasizing the substantial strain imposed by satellite manufacturing on global mineral resources \cite{kumaran_quantifying_2024}. Pardini and Anselmo introduced environmental criticality indexes to assess the sustainability of space activities, particularly in LEO. Their findings reveal that between one-third and one-half of LEO's capacity to sustain long-term space activities has already been saturated, underscoring the need for immediate action to manage this finite resource \cite{pardini_environmental_2020, pardini_evaluating_2021}.

Future projections suggest a substantial increase in the environmental impact of the space sector. A comprehensive LCA conducted by \cite{miraux_environmental_2022-1} found that by 2050, proposed large constellations and other space activities could result in a ninefold increase in the sector’s climate change impacts compared to 2021 baseline activities, rising from 4.1 million tons to 760 million tons CO$_2$ annually. However, since life cycle impacts must be evaluated holistically rather than focusing solely on individual stages, optimizing vehicle materials and manufacturing methods is essential to achieving meaningful reductions in these emissions over their full operational lifespan \cite{kirchain_jr_environmental_2017}. For instance, harnessing nonlinear orbital perturbation forces can reduce propellant and maintenance costs, potentially decreasing the vehicle mass budget for propellant by approximately 60\% \cite{singh_low_2020}. Additionally, innovative propulsion systems or hybrid rocket propulsion systems could further enhance sustainability \cite{barato_review_2023}.

Initial assessments of emissions have primarily focused on American \cite{noauthor_space_nodate} and European launch vehicles \cite{maury_application_2020}. However, our work highlights the critical need to incorporate a broader range of international launch vehicles into environmental analyses, as this represents a significant step forward in achieving a more comprehensive understanding of the global impact of satellite constellations. As the scope of satellite constellations grows to involve an order-of-magnitude more satellites, this increasing scale underscores the importance of analyzing heavy-lift rockets developed by space-faring nations such as China (Long March) and India (LVM3), which could play significant roles in future deployments.

Each system has a unique emission profile and operational characteristics that must be factored into environmental appraisals.

Satellite megaconstellations offer significant societal benefits, including global broadband connectivity and advancements in communication technologies. However, their rapid growth also presents profound challenges to environmental and space sustainability. These challenges demand a balanced and holistic approach to environmental stewardship that extends to the orbital domain. 

Efforts to quantify the long-term impacts and promote sustainable practices are essential. Advancing reusability in launch vehicles, optimizing satellite lifecycles, and incorporating eco-friendly materials and propellants represent promising pathways forward. As the space sector evolves, maintaining a balance between innovation and sustainability will be key to its long-term viability.

\section{Methodology}\label{sec4}

For the purposes of this analysis, constellations with at least 150 planned satellites were categorized as megaconstellations. All such constellations with ITU filings are included in the emissions estimates presented here \cite{noauthor_e-submission_nodate}, with the exception of the SpinLaunch constellation. Unlike traditional systems, SpinLaunch utilizes a kinetic launch mechanism rather than rockets \cite{noauthor_spinlaunch_nodate}, and due to the lack of publicly available data on its emissions, it was excluded from this study.

Our methodology involves conducting a comprehensive LCA of emissions associated with satellite megaconstellations. This includes detailed calculations of emissions from multiple stages of rocket launches and satellite operations, while incorporating the effects of reusability. The stages considered in our analysis encompass:

\begin{enumerate}
    \item \textbf{Propellant Combustion} – Emissions released during rocket launches.
    \item \textbf{Rocket Launcher Production} – Environmental costs of manufacturing launch vehicles.
    \item \textbf{Electricity Consumption} – Energy usage for ground operations and satellite support.
    \item \textbf{Transportation} – Emissions from logistics and transportation of launch vehicles and components.
    \item \textbf{Satellite Production} – Environmental impacts of building and assembling satellites.
\end{enumerate}

In addition to estimating the overall emissions from these stages, we assess their environmental impact on a per-subscriber basis under various scenarios, providing a perspective on the emissions intensity relative to the service provided.

The following equations form the foundation of our emission analysis, providing a mathematical framework for quantifying and comparing emissions across different constellations.

\subsection{Total Emissions from Propellant Combustion}

Propellant combustion is the primary source of emissions in rocket launches. The total emissions from propellant combustion \( E_{\text{propellant}} \) are calculated using the following equation:

\begin{equation}
E_{\text{propellant}} = \sum_{i=1}^{N} \left( M_{i} \times EF_{i} \right)
\end{equation}

The term \( M_{i} \) denotes the mass of the \( i^{\text{th}} \) propellant type, while \( EF_{i} \) is the corresponding emission factor, indicating the amount of emissions per unit mass of propellant. The summation is performed across all \( N \) propellant types used in the analysis.

In this study, the analysis includes four common propellant types:
\begin{itemize}
    \item \textbf{Solid} – Ammonium perchlorate (NH\(_4\)ClO\(_4\)), aluminum (Al), and hydroxyl-terminated polybutadiene (HTPB),
    \item \textbf{Cryogenic} – Liquid oxygen (LO\(_x\)), liquid hydrogen (LH\(_2\)), and water vapor (H\(_2\)O),
    \item \textbf{Kerosene-based} – Liquid oxygen (LO\(_x\)), Refined Petroleum-1 (RP-1), and methane (CH\(_4\)),
    \item \textbf{Hypergolic} – Nitrogen tetroxide (N\(_2\)O\(_4\)) and unsymmetrical dimethylhydrazine (UDMH).
\end{itemize}

Emission factors (\( EF_{i} \)) for these propellants were sourced from Ross et al. \cite{ross_limits_2009}. Since most modern rockets utilize a combination of these propellants, the total emissions calculation accounts for the weighted contributions of each type. This approach ensures an accurate representation of the emissions profile for mixed-propellant systems.

\subsection{Total Production Emissions}

The production of rockets contributes significantly to emissions, arising from the materials and electronics used in their construction. The total emissions from rocket production are calculated using the following equation:

\begin{equation}
E_{\text{production}} = DM_{j} \times EF_{j} + DM_{e} \times EF_{e}
\end{equation}
\\
In this equation, \( E_{\text{production}} \) represents the total emissions from rocket production. The term \( DM_{j} \) denotes the dry mass of material \( j \), with \( EF_{j} \) as the corresponding emission factor for material \( j \). Similarly, \( DM_{e} \) refers to the dry mass of the electronics components, and \( EF_{e} \) is the emission factor associated with electronics manufacturing.

For this analysis, all rockets except SpaceX's Starship are assumed to be constructed using an aluminum-lithium alloy, which consists of approximately 97\% aluminum and 3\% lithium. Starship, on the other hand, is expected to use stainless steel, and the respective emission factors for these materials have been applied. These factors were sourced from reliable material databases to ensure precision in calculations. \cite{noauthor_international_nodate}

Additionally, the dry mass of the electronics is estimated to be approximately 5\% of the total dry mass of the rocket. This figure is based on historical data from NASA vehicle designs and is used consistently across the analysis to estimate the electronics-related emissions \cite{motiwala_conceptual_2014}. This category encompasses various electronic components essential for the functionality and control of the rocket, including avionics systems, flight computers, sensors, communication systems, power distribution units, telemetry systems, and guidance, navigation, and control (GNC) hardware. By accounting for both the structural materials and the electronics, this method provides a comprehensive estimate of the production emissions for each rocket design.

\subsection{Electricity Emissions}

The production of rocket components requires significant electricity, which contributes to the overall emissions footprint. The total emissions from electricity consumption are calculated using the following equation:

\begin{equation}
E_{\text{electricity}} = DM_{j} \times C_{j} \times E_{j}
\end{equation}
\\
In this equation, \( E_{\text{electricity}} \) represents the total emissions resulting from electricity consumption during production. The term \( DM_{j} \) denotes the dry mass of material \( j \), while \( C_{j} \) is the energy consumption required per kilogram of material \( j \). The variable \( E_{j} \) represents the emission factor per kilowatt-hour (kWh) of electricity used in the production process.

This formulation allows for the estimation of emissions based on the specific material composition of each rocket and the energy-intensive processes involved in their production. The emission factor \( E_{j} \) is derived from the US Energy Information Administration (EIA) for rockets manufactured in the US \cite{barato_investigation_2024} and from the International Energy Agency (IEA) for all other rockets \cite{noauthor_emissions_nodate}, to adjust for country-specific energy mixes and associated emissions intensities. This approach ensures that the model accounts for regional variations in energy generation and their respective environmental impacts.

\subsection{Adjusted Production and Electricity Emissions for Reusable Rockets}

Reusable rockets have a distinct emissions profile due to their ability to reduce the need for new production and electricity for each subsequent launch. The adjusted production and electricity emissions for reusable rockets are calculated as:

\begin{equation}
E_{\text{reusable}} = \frac{E_{\text{production}} + E_{\text{electricity}}}{RU}
\end{equation}
\\
In this equation, \( E_{\text{reusable}} \) denotes the adjusted emissions per launch, and \( RU \), the reusability factor, accounts for the reduction in emissions due to reuse. The reusability factor is defined as:

\begin{equation}
RU = S1P \times R \times (1 - R_A)^{\left(\frac{R}{L}\right)}
\end{equation}
\\
Here, \( S1P \) represents the proportion of the rocket's mass attributed to the first stage, while \( R \) is the number of times the stage is reused. The refurbishment factor, \( R_A \), indicates the proportion of the rocket requiring refurbishment after each reuse cycle and is assumed to be a constant value of 0.1 (or 10\%) for all rocket types, including any emissions from transportation to refurbishment facilities. The parameter \( L \) represents the number of launches before a major refurbishment is required, influencing the extent of reuse efficiency.

This equation is applied to partially reusable rockets such as Falcon 9, Falcon Heavy, and New Glenn, where only Stage 1 is reused. For fully reusable rockets like SpaceX's Starship, the formula is modified to account for the reusability of the entire rocket. The reusability factor for Starship is expressed as:

\begin{equation}
RU = R \times (1 - R_A)^{\left(\frac{R}{L}\right)}
\end{equation}
\\
Unlike partially reusable rockets, the \( S1P \) term is omitted for Starship, as its entire structure is designed for reusability. This distinction reflects the advanced design of fully reusable systems, which offer a greater potential for emissions reduction. By incorporating these calculations, the model captures the environmental benefits of reusability while enabling a fair comparison between reusable and expendable rockets. Thus, this factor allows for production emissions to be distributed over the lifetime of the asset.

\subsection{Transportation Emissions}

The transportation of rocket components to launch sites contributes significantly to the total emissions associated with rocket launches. These emissions are calculated using the following equation:

\begin{equation}
E_{\text{transport}} = \sum_{v=1}^{V} \left( M_{v} \times D_{v} \times EF_{v} \right)
\end{equation}
\\
In this equation, \( E_{\text{transport}} \) represents the total emissions from transportation. The term \( M_{v} \) denotes the mass transported by vehicle type \( v \), \( D_{v} \) is the distance traveled by that vehicle type, and \( EF_{v} \) is the emission factor for the corresponding vehicle type. The summation is performed across \( V \), the total number of distinct vehicle types involved in transportation.

Emission factors were considered for four primary vehicle types used in transporting rocket components: trucks, trains, cargo aircraft, and container ships \cite{congressional_budget_office_emissions_nodate}. Notably, for the Ariane 6 rocket, the emission factor for container ships was adjusted to account for the use of the \textit{Canopée}, an eco-friendly container ship equipped with sails that reduce emissions by approximately 30\% \cite{werner_orcelle_2023}. The \textit{Canopée}, in operation since 2022, reflects the unique efforts made to minimize the environmental impact of transportation for specific rockets.

The distance traveled by each vehicle type was calculated as the shortest possible route between the rocket’s manufacturing site and its launch site. For ground transport, distances were determined using Google Maps, while air and sea routes were estimated using Google Earth. This method ensures consistency and accuracy in estimating transportation-related emissions.

\subsection{Launch Emissions Per Rocket}

The total emissions for each rocket launch, encompassing adjusted production emissions, propellant combustion, and transportation emissions, are calculated as:

\begin{equation}
E_{\text{launch\_per\_rocket}} = E_{\text{reusable}} + E_{\text{propellant}} + E_{\text{transport}}
\end{equation}
\\
This equation provides a comprehensive measure of the emissions associated with a single rocket launch, combining the key contributors to its environmental impact. This enables direct comparisons between different launch vehicles by providing a standardized measure of their total environmental impact per launch.

\subsection{Total Launch Emissions for Satellite Constellations}

The total emissions from launches required to deploy satellite constellations are given by:

\begin{equation}
E_{\text{launch}} = \sum_{m=1}^{R} \left( \frac{N_{\text{satellite},m}}{N_{\text{per\_rocket},m}} \times E_{\text{launch\_per\_rocket}} \right)
\end{equation}
\\
In this equation, \( E_{\text{launch}} \) represents the total emissions from all launches across the constellations analyzed. The variable \( N_{\text{satellite},m} \) denotes the total number of satellites in constellation \( m \), while \( N_{\text{per\_rocket},m} \) is the number of satellites carried per rocket for that constellation. The summation is performed across \( R \), the total number of constellations included in the analysis.

To conduct this LCA, certain assumptions were necessary to address gaps in publicly available data. For constellations that utilize multiple rocket types, the most frequently used rocket was selected for emissions calculations to ensure consistency. In cases where no information was available on the number of satellites per rocket, this value was assumed to be 10, based on typical payload capacities. For constellations employing rockets manufactured in China, such as the Long March, Kuaizhou-1, or Zhuque-2, the Long March 5 rocket was assumed to be representative. This decision reflects the limited emissions data available for other Chinese rocket models and the expected similarity in their emissions profiles.

This approach enables a standardized approach to emissions estimation across a diverse range of constellations while ensuring methodological consistency and transparency. We acknowledge the inherent uncertainty introduced by these assumptions and will address this uncertainty through the Uncertainty Quantification (UQ) method, ensuring that our results remain robust and reliable despite potential data limitations.

\subsection{Total Emissions from Satellite Production}

The production of satellites contributes to overall emissions, primarily due to the manufacturing of krypton and xenon, which are extensively used in ion propulsion systems. The total emissions from satellite production are calculated as:

\begin{equation}
E_{\text{satellite}} = \sum_{n=1}^{S} \left( M_{\text{sat},n} \times \left( X_{\text{kx}} \times E_{\text{kx}} \right) \right)
\end{equation}

In this equation, \( E_{\text{satellite}} \) represents the total emissions resulting from satellite production. The term \( M_{\text{sat},n} \) denotes the mass of satellite \( n \), while \( X_{\text{kx}} \) is the proportion of krypton or xenon within the satellite mass. For the purposes of this analysis, \( X_{\text{kx}} \) is assumed to be a constant value of 2\% across all satellites. The variable \( E_{\text{kx}} \) represents the emission factor for the production of krypton and xenon, reflecting the environmental impact of their manufacturing processes. The summation is performed across all satellites in the constellation, as indicated by \( S \).

This approach ensures a standardized estimate of emissions for satellite production, highlighting the significance of propulsion system components in the overall environmental footprint. The constant proportion of krypton and xenon provides a consistent basis for comparison, while the emission factor, sourced from \cite{ross_limits_2009}, incorporates the specific impacts of producing these critical elements.

\subsection{Number of Subscribers}

The number of subscribers per satellite is calculated using the following equation:

\begin{equation}
N_{\text{subscriber}} = \frac{N_{\text{satellite}} \times S_{\text{factor}}}{N_{\text{constellation}}}
\end{equation}
\\
In this equation, \( N_{\text{subscriber}} \) represents the number of subscribers per satellite. The variable \( N_{\text{satellite}} \) is the total number of satellites in the constellation, while \( S_{\text{factor}} \) denotes the subscriber factor, defined as the average number of subscribers supported per satellite. The term \( N_{\text{constellation}} \) indicates the total number of constellations under analysis.

The subscriber factor \( S_{\text{factor}} \) was determined using publicly available data from the Starlink constellation \cite{ma_network_2023}, as it is the only megaconstellation analyzed that has released information on its current number of subscribers. For other constellations, where subscriber data is not available, \( S_{\text{factor}} \) was assumed to be similar to that of Starlink, regardless of the total number of planned satellites. While this assumption ensures consistency across the analysis, we explicitly account for the variability among constellations—arising from differences in operational goals, target markets, and technology—through the UQ analysis. By incorporating these uncertainties into our Monte Carlo framework, we ensure that our results remain robust despite potential discrepancies in subscriber estimates.

\subsection{Per Subscriber Emissions}

The emissions attributable to each subscriber are calculated using the equation:

\begin{equation}
E_{\text{subscriber}} = \frac{E_{\text{launch}} + E_{\text{satellite}}}{N_{\text{subscriber}}}
\end{equation}

In this equation, \( E_{\text{subscriber}} \) represents the emissions per subscriber. The numerator combines the total emissions from rocket launches, \( E_{\text{launch}} \), and satellite production, \( E_{\text{satellite}} \), capturing the overall environmental impact of deploying and maintaining the satellite constellation. The denominator, \( N_{\text{subscriber}} \), represents the number of subscribers per satellite, as determined in the previous section.

It is also important to note that competition among constellations for a limited subscriber market could lead to fewer subscribers per satellite than projected. This would further increase the emissions per subscriber, highlighting the environmental risks associated with market overestimation.

This formulation provides a per-subscriber metric that reflects the environmental cost of satellite-based communication networks. By incorporating both launch and production emissions, the calculation ensures a comprehensive assessment of the emissions intensity associated with satellite services.

\subsection{Uncertainty Quantification}
To systematically account for uncertainty in our calculations, we employ a Monte Carlo approach with a Gaussian distribution. Each variable is modeled with a mean value derived from the calculations in the previous sections and a standard deviation of \( 0.25 \  \times \) the mean value, ensuring a realistic range of variation across different constellations. This results in lower-bound and upper-bound estimates, corresponding to a \( \pm 25\% \) deviation from the mean.

By repeatedly sampling from this distribution, the model generates a probability distribution for each variable rather than relying on fixed deterministic values. This probabilistic approach improves robustness by capturing real-world fluctuations in key parameters, such as satellite production, launch emissions, and transportation factors. Consequently, emissions estimates also follow a distribution, reflecting the inherent variability in these processes.

The Monte Carlo simulation runs thousands of iterations, enabling a comprehensive sensitivity analysis that highlights the full spectrum of potential emissions outcomes under varying conditions. This methodology not only strengthens confidence in the modeled results but also provides a more nuanced understanding of emissions intensity, facilitating more informed decision-making for policymakers and industry stakeholders.

\subsection{Rocket Specifications}

\begin{figure}[h]
    \centering
    \includegraphics[width=0.95\textwidth]{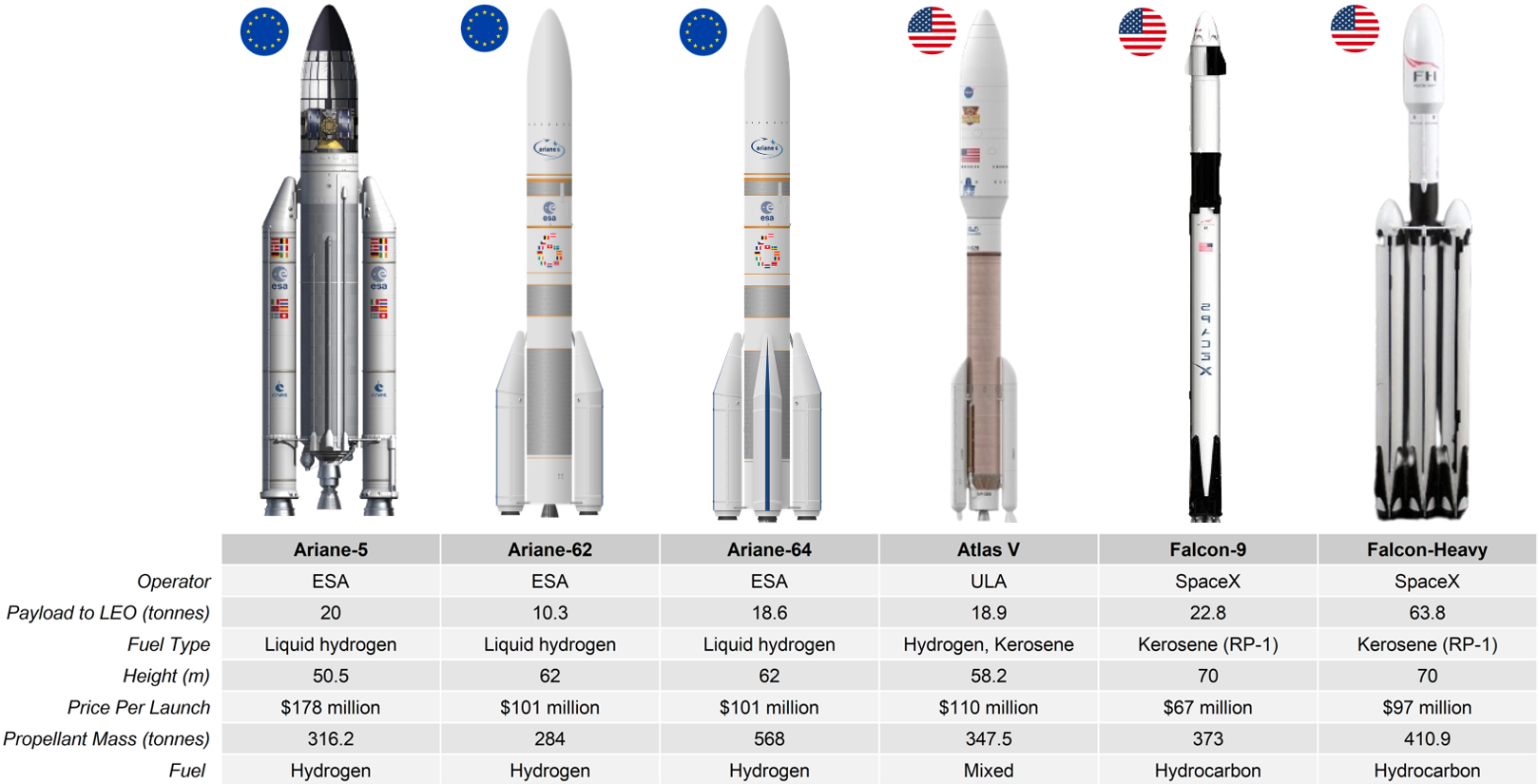}
    \includegraphics[width=0.95\textwidth]{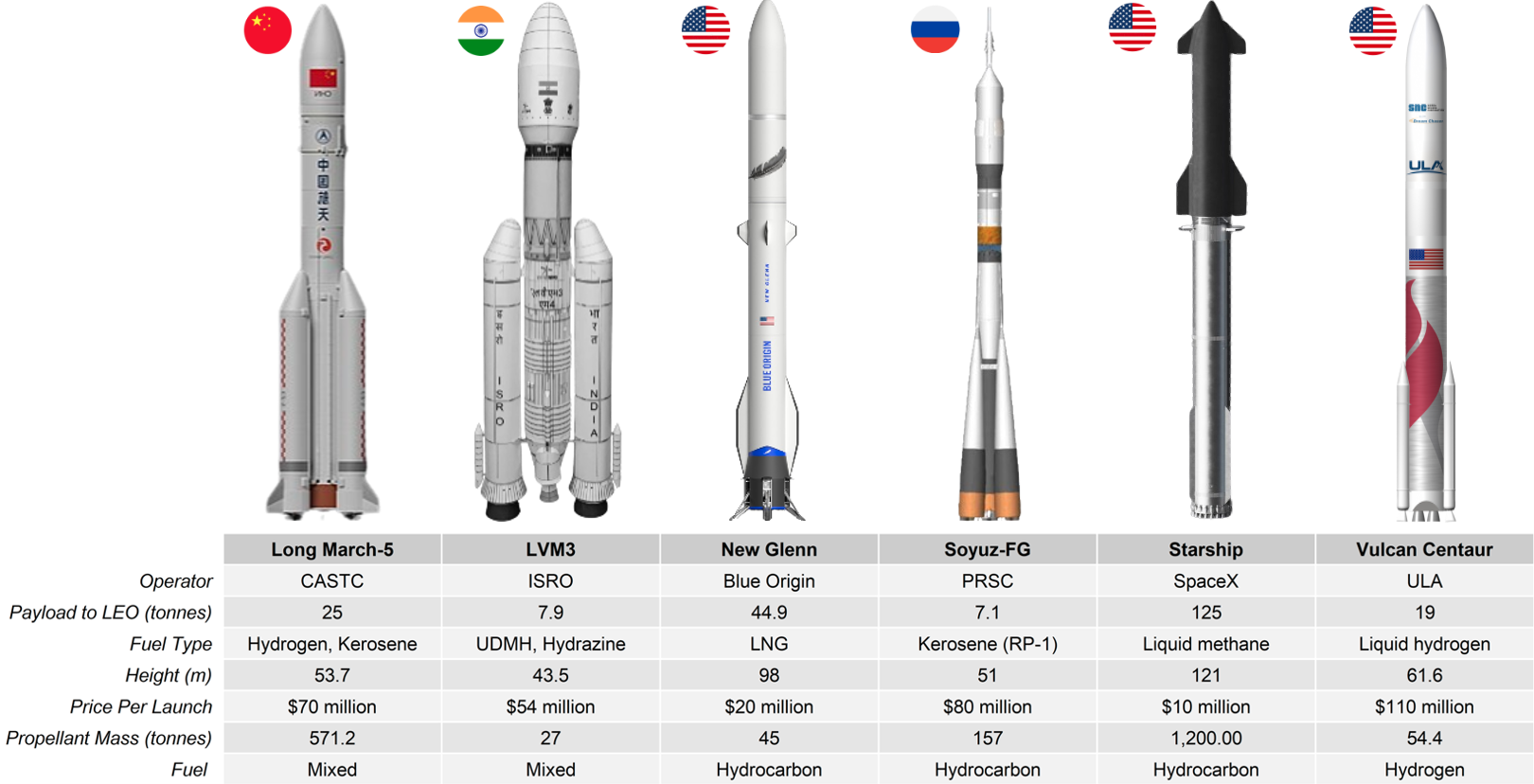}
    \caption{Specifications of the rockets commonly used to launch satellite constellations, highlighting key metrics such as payload capacity, fuel type, and propellant mass.}
    \label{fig:rockets}
\end{figure}

Figure \ref{fig:rockets} provides a detailed overview of 12 rockets commonly used for launching satellite constellations, utilizing data from the rocket manufacturer websites \cite{noauthor_esa_nodate-1, noauthor_esa_nodate, noauthor_atlas_nodate, noauthor_spacex_nodate-2, noauthor_spacex_nodate-1, noauthor_cnsa_nodate, noauthor_indian_nodate, noauthor_new_nodate, noauthor_facts_nodate, noauthor_spacex_nodate, noauthor_vulcan_nodate}. The figure presents key specifications critical for conducting an LCA, including payload capacity, fuel type, and propellant mass. These rockets are operated by space organizations and corporations from across the globe, including Europe (ESA), United States (SpaceX, ULA, Blue Origin), China (CASTC), India (ISRO), and Russia (PRSC).

The rockets utilize a variety of fuels, including liquid hydrogen, kerosene (RP-1), UDMH, hydrazine, liquid natural gas (LNG), and liquid methane, reflecting the diversity of propulsion technologies. Payload capacities range significantly, from 7.1 tonnes (Soyuz-FG) to 125 tonnes (Starship), highlighting the varying scales of missions they support. Propellant mass also varies substantially, spanning from as low as 27 tonnes (LVM3) to 1,200 tonnes (Starship), illustrating the heterogeneity in fuel requirements for different rocket designs and mission profiles.

The price per launch further emphasizes the variability among these rockets, with costs ranging from \$10 million (New Glenn) to \$178 million (Ariane-5), reflecting differences in technology, capacity, and operational efficiency. Such variations underline the importance of including these specifications in the LCA to account for the environmental impact and economic considerations of satellite deployment.

The figure also highlights the international geostrategic competition underlying rocket development, as space launch capabilities are closely tied to national sovereignty and technological leadership. This comprehensive dataset serves as the foundation for analyzing the environmental impacts of the rockets studied in this work.

\subsection{Parameters}

\begin{figure}[h]
    \centering
    \includegraphics[width=0.95\textwidth]{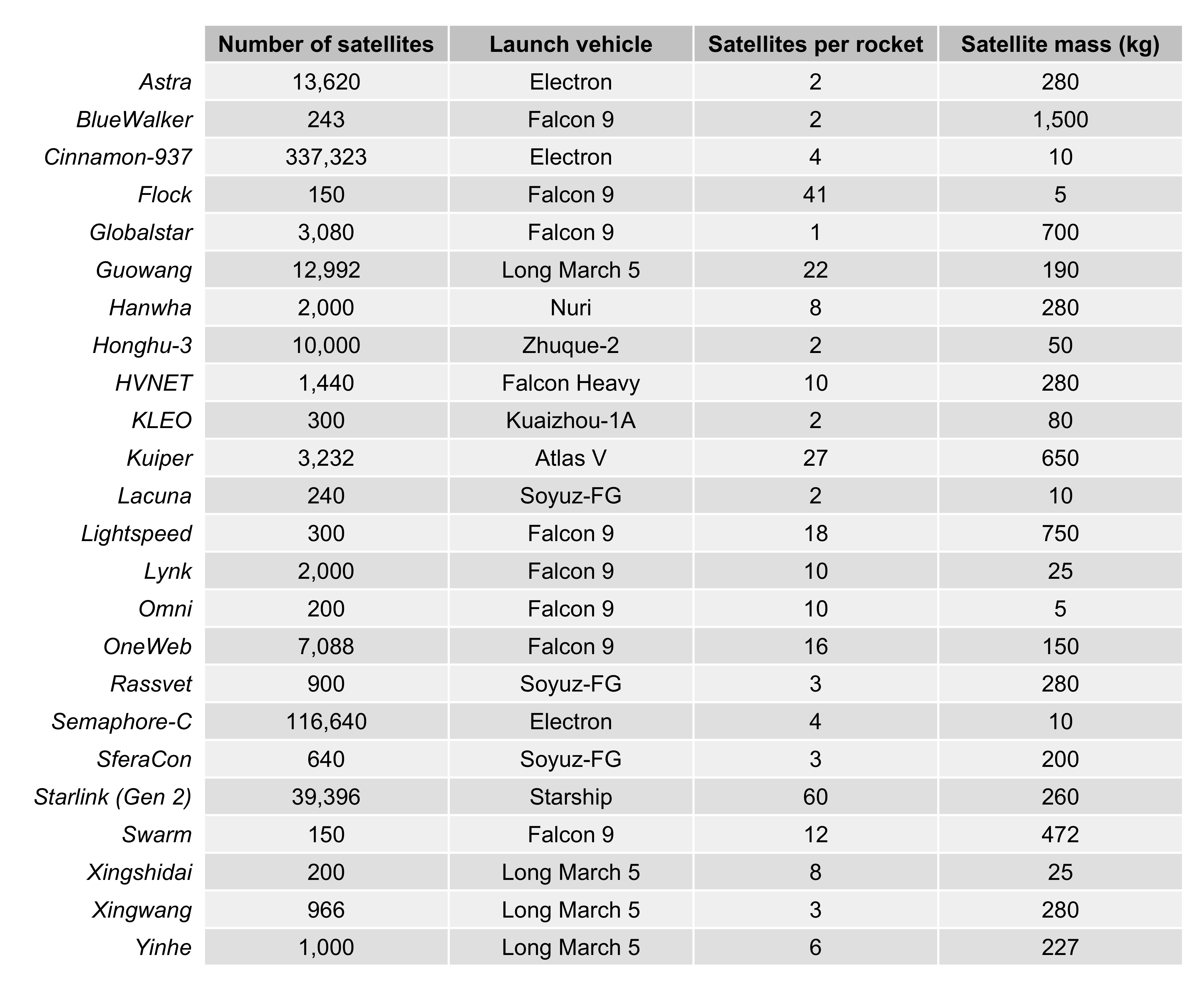}
    \caption{Key parameters for the satellite constellations analyzed in the LCA.}
    \label{fig:constellation_parameters_table}
\end{figure}

\begin{figure}[h]
    \centering
    \includegraphics[width=0.95\textwidth]{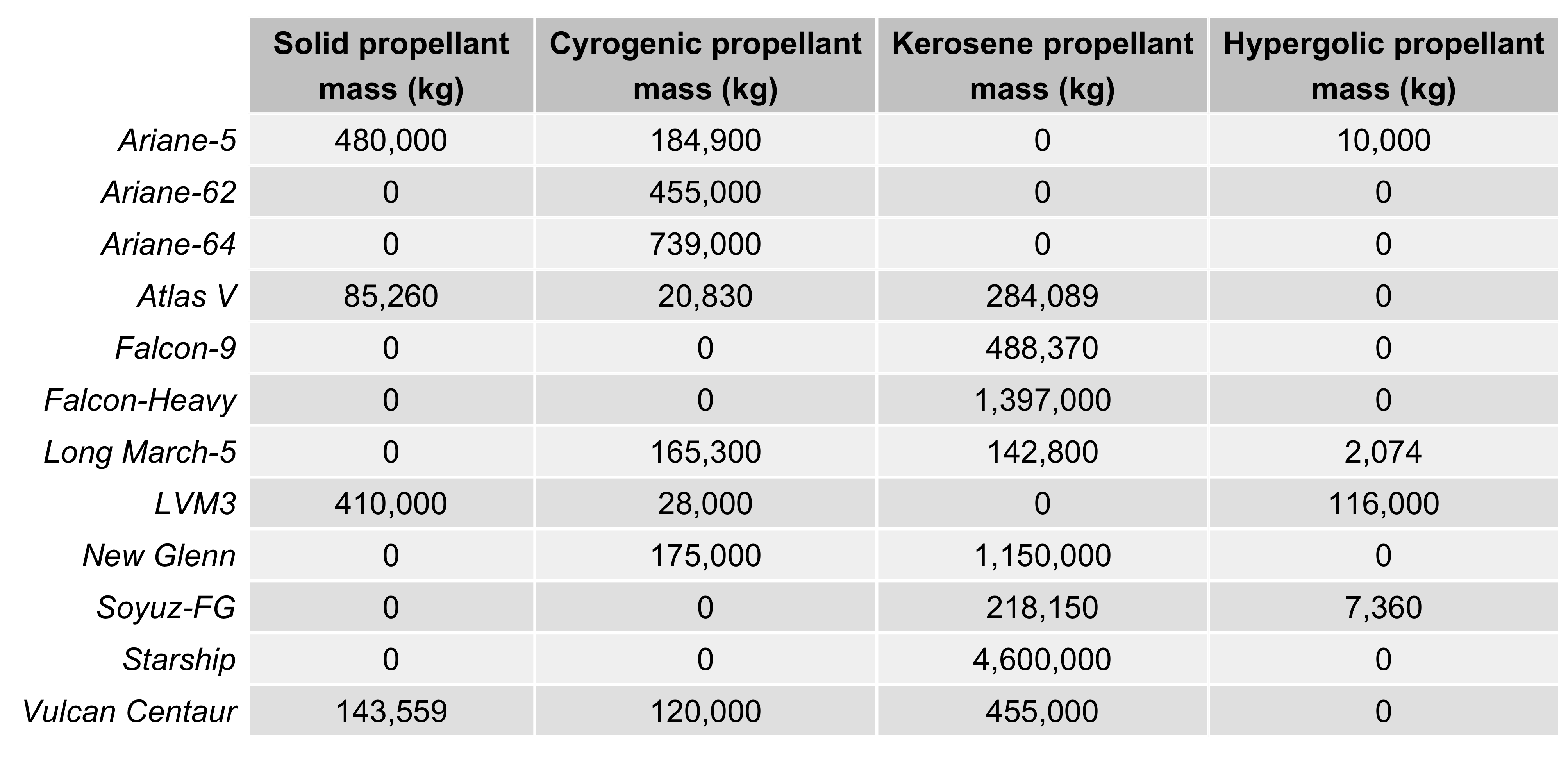}
    \caption{Propellant-specific parameters for the launch vehicles analyzed in the LCA.}
    \captionsetup{labelformat=default,labelsep=period,name=Table}
    \label{fig:rocket_parameters_table_1}
\end{figure}

\begin{figure}[h]
    \centering
    \includegraphics[width=0.95\textwidth]{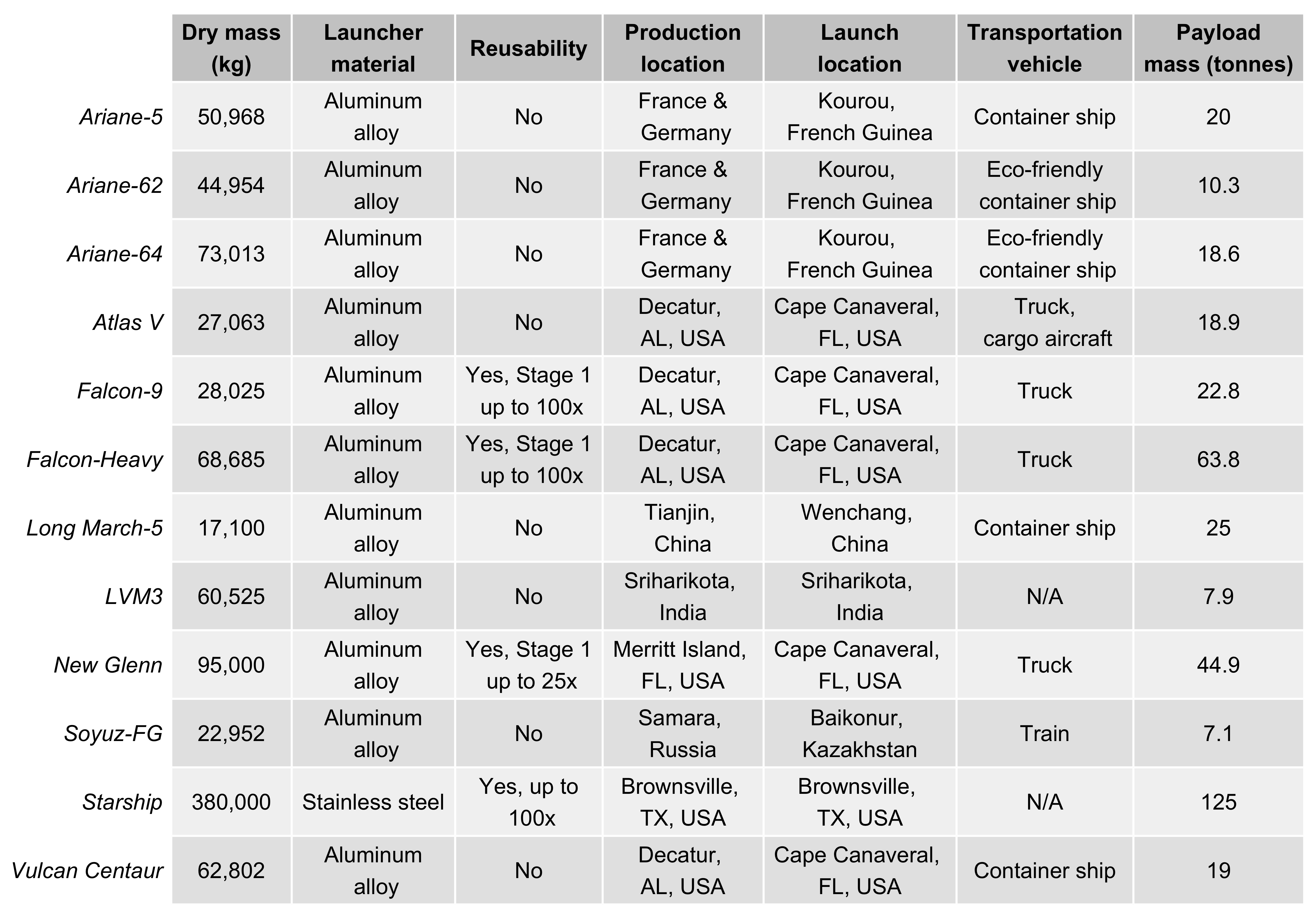}
    \caption{Additional parameters for the launch vehicles analyzed in the LCA.}
    \captionsetup{labelformat=default,labelsep=period,name=Table}
    \label{fig:rocket_parameters_table_2}
\end{figure}

To conduct a comprehensive LCA of satellite megaconstellations, we utilized a detailed set of parameters for both satellite constellations and the launch vehicles, utilizing data from ITU filings and rocket manufacturer websites. These parameters, summarized in Figures  \ref{fig:constellation_parameters_table}, \ref{fig:rocket_parameters_table_1}, and \ref{fig:rocket_parameters_table_2}, include critical data related to satellite deployment, launch vehicle specifications, and emissions factors, forming the basis for calculating the environmental impacts.

Figure \ref{fig:constellation_parameters_table} outlines the parameters for the satellite constellations, including the number of satellites, the associated launch vehicle, satellites deployed per rocket, and the satellite mass (kg). These details are critical for modeling emissions across the full lifecycle of satellite constellations. For constellations where launch vehicles or other parameters are still uncertain, we made our best estimates based on the available data.

Figures \ref{fig:rocket_parameters_table_1} and \ref{fig:rocket_parameters_table_2} present the launch vehicle parameters, covering propellant types and quantities, dry mass, launcher materials, reusability, production and launch locations, transportation methods, and payload capacities. These parameters are integral to quantifying emissions during the launch phase, one of the most significant contributors to the total GHG emissions for satellite constellations. For instance, rockets like Falcon 9 and Falcon Heavy utilize kerosene-based fuels, while Starship employs methane, reflecting variations in emissions intensity across different propulsion technologies.

The data also highlight the trade-offs between reusability and emissions. Rockets such as Falcon 9, Falcon Heavy, and Starship, which feature high reusability, can significantly reduce per-launch emissions compared to non-reusable rockets like the Long March 5 and Soyuz-FG. Additionally, the production and transportation parameters reveal the impact of logistics on overall emissions, with rockets transported by eco-friendly container ships (e.g., Ariane-62 and Ariane-64) having a reduced transportation emissions profile.

\begin{figure}[h]
    \centering
    \includegraphics[width=0.95\textwidth]{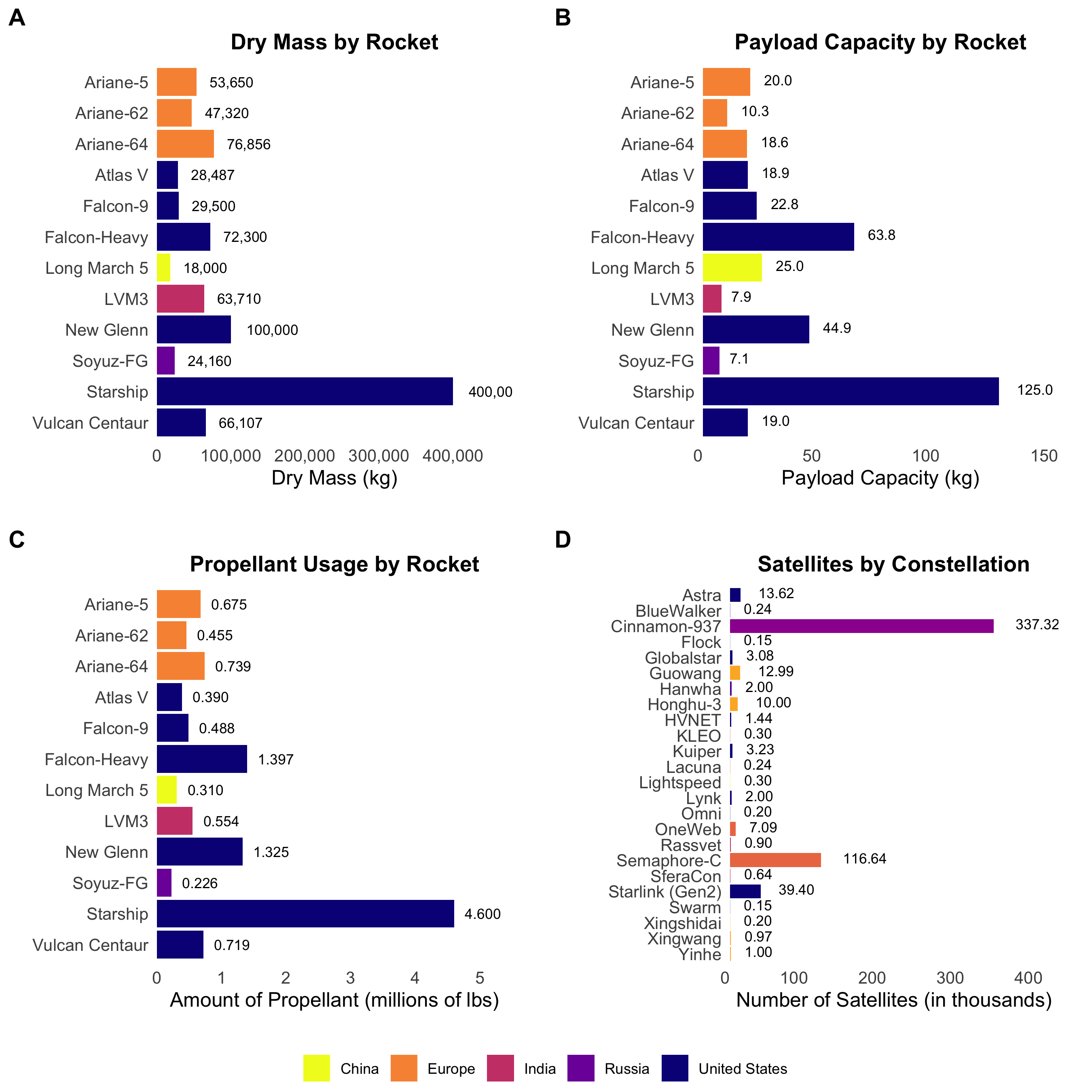}
    \caption{Parameters for emissions calculations for launch vehicles and constellations.}
    \label{fig:parameters_and_constants}
\end{figure}

Figure \ref{fig:parameters_and_constants} further details the constants and emissions factors used in the calculations, including material-specific factors for aluminum alloy, steel, and composites, as well as fuel types such as cryogenic, kerosene, and hypergolic propellants. For example, aluminum alloy, a common material in launcher production, has an emissions factor of 14.9~kg CO$_2$ per kg, while steel used in Starship exhibits a lower emissions factor of 6.6~kg CO$_2$ per kg. Propellant-specific emissions factors, such as 3.15~kg CO$_2$ per kg for kerosene and 0.89~kg CO$_2$ per kg for liquid hydrogen, enable accurate modeling of combustion-related emissions.

The figure also includes transportation-related emissions factors, with cargo ships producing 0.72~kg CO$_2$ per kg of payload transported, compared to trucks and planes, which have factors of 0.03--0.11~kg CO$_2$ per kg. These factors highlight the logistical considerations that influence overall emissions, especially for rockets requiring intercontinental transport of components.

Together, these figures provide a comprehensive dataset for emissions modeling, capturing the diverse range of inputs influencing the environmental impact of satellite megaconstellations. This parameterization allows for detailed analyses, enabling comparisons between deployment strategies, materials, reusability levels, and propellant choices, ensuring robust and accurate LCA results.

\subsection{Methodology Overview and Code Accessibility}

The equations and parameters outlined in this study enable a systematic analysis of the environmental impact of satellite megaconstellations, offering critical insights into their sustainability and emissions profiles. This framework facilitates a comprehensive LCA that accounts for production, transportation, launch, and operational phases.

The code used for this analysis is part of the Open-source Rocket and Constellation Lifecycle Emissions (ORACLE) project \cite{kukreja_open-source_nodate}. The repository includes visualization scripts developed in R, along with the input data required to replicate and expand upon the results presented in this study. By making this code accessible, the ORACLE project encourages transparency and fosters collaboration in the study of space sustainability.

\section{Results}\label{sec5}

The results of our study provide a comprehensive overview of the GHG emissions associated with satellite megaconstellations, focusing on different stages of rocket launches and satellite operations. 

\subsection{Emissions Per Rocket}

\begin{figure}[h]
    \centering
    \includegraphics[width=0.95\textwidth]{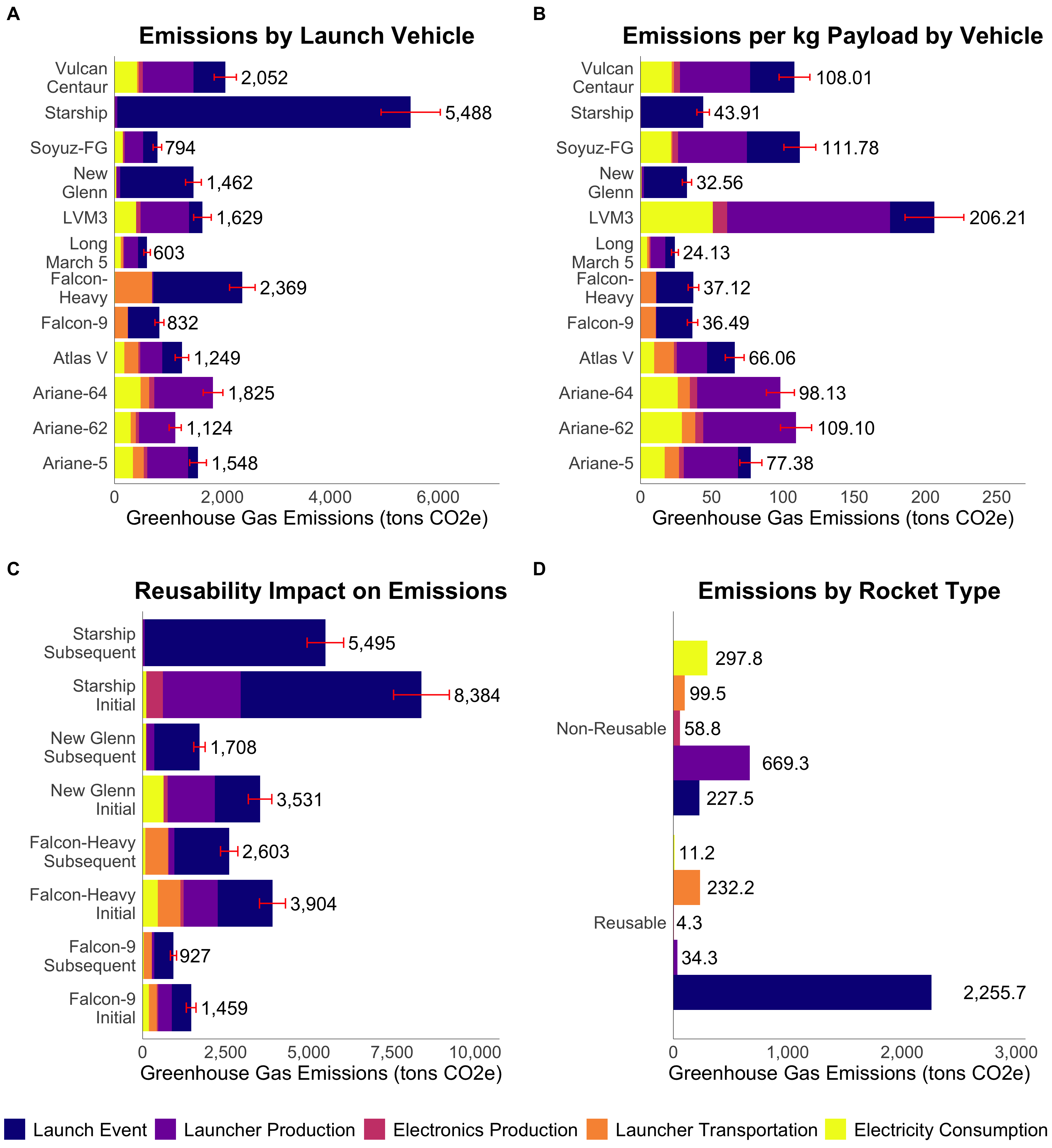}
    \caption{Total and adjusted GHG emissions for each launch vehicle.}
    \label{fig:rocket_emissions}
\end{figure}

Figure \ref{fig:rocket_emissions} provides a comprehensive analysis of GHG emissions associated with various launch vehicles, highlighting total emissions, payload-adjusted emissions, and the effects of reusability on their environmental impact.

The average total emissions for a rocket are approximately 1.75 kt per launch, equivalent to the annual carbon absorption of 80,000 trees \cite{noauthor_calculation_nodate}. The breakdown of these emissions is as follows: 903 t from propellant combustion (51.6\%), 458 t from launcher production (26.2\%), 41 t from electronics production (2.3\%), 144 t from transportation (8.2\%), and 202 t from electricity consumption (11.5\%).

It is also important to note that some rockets, such as the Ariane 62 and the Ariane 64 do not have any CO$_2$ emissions from the launch event due to their use of only cryogenic propellant.

As shown in Figure \ref{fig:rocket_emissions}A, the Starship rocket (with a payload of 125 t) exhibits the highest emissions, producing approximately 5.49 kt of CO$_2$ per launch, with 99\% of these emissions stemming from propellant combustion during the launch event. In contrast, the Long March 5 (with a payload of 25 t) demonstrates the lowest emissions at 603 t per launch, while maintaining a similar distribution of emissions among its stages as other rockets. Across all rockets, the launch event consistently dominates as the primary source of emissions (51.6\%), while launcher production (26.2\%), electronics production (2.3\%), transportation (8.2\%), and electricity consumption (11.6\%) play smaller roles.

When emissions are adjusted to account for payload masses, the relative efficiency of the rockets (pictured in Figure \ref{fig:rocket_emissions}B) shifts significantly. LVM3 stands out with the highest adjusted emissions, reaching 206 t of CO$_2$ per kilogram of payload mass—well above the average of less than 80 t. On the other hand, the Long March 5 retains the lowest adjusted emissions, emphasizing its efficiency even after normalization. This adjustment highlights the importance of payload capacity as a critical factor in evaluating the environmental impact of launch systems.

Reusable rockets, such as Falcon 9, Falcon Heavy, and New Glenn, demonstrate substantially lower production emissions compared to their non-reusable counterparts, as evident from Figure \ref{fig:rocket_emissions}D. For these rockets, production emissions account for only 1.05\%, 0.90\%, and 4.50\% of their total emissions, respectively, compared to an average of 26.2\% across all analyzed launch vehicles. Moreover, these reusable rockets achieve the lowest emissions per kilogram of payload mass, with all of them producing less than 38,000~kg of CO$_2$ per kilogram of payload. Notably, some rockets, such as LVM3, New Glenn, Soyuz-FG, and Starship, exhibit minimal emissions from launcher transportation, contributing less than 0.7\% of their total emissions—significantly below the average of 8.91\%.

The impact of reusability on emissions is further highlighted in comparisons between initial and subsequent launches in Figure \ref{fig:rocket_emissions}C. For reusable rockets, subsequent launches result in significantly reduced emissions. For instance, Falcon 9’s emissions decrease from 1,459 t during an initial launch to 819 t for subsequent launches. This trend underscores the environmental benefits of reusability in reducing the overall emissions associated with space exploration.

\subsection{Emissions Per Satellite Constellation}

\begin{figure}[h]
    \centering
    \includegraphics[width=0.95\textwidth]{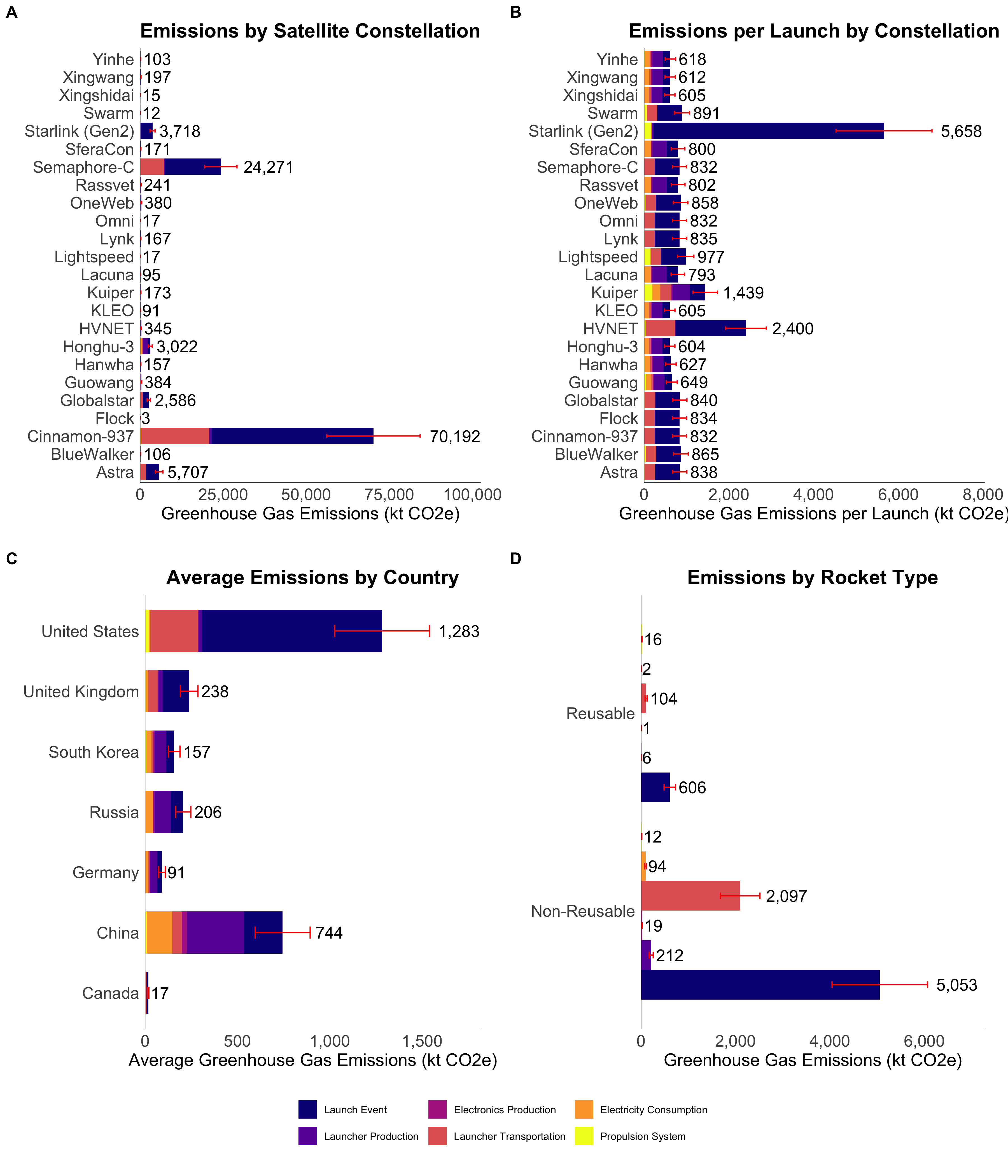}
    \caption{Total and adjusted GHG emissions for each satellite megaconstellation \& emissions for megaconstellations by country and by rocket type.}
    \label{fig:constellation_emissions}
\end{figure}

The GHG emissions associated with satellite megaconstellations reveal significant variability across deployment strategies and constellation sizes. Figure~\ref{fig:constellation_emissions} illustrates the total emissions from rocket launches, satellite production (including ion propulsion systems), and other associated processes.

Notably, as shown in Figure~\ref{fig:constellation_emissions}A, the Cinnamon-937 constellation exhibits the highest total emissions, surpassing 70,000 kt, while the Flock constellation records the lowest emissions at approximately 3 kt. These disparities highlight the dominant role of larger constellations in driving cumulative environmental impacts.

Across all constellations, the majority of emissions (95.6\%) originate from two primary sources: the launch event (68.5\%) and launcher transportation (27.1\%). Other stages, such as electronics production and satellite propulsion systems, contribute comparatively minor fractions to the total footprint. For instance, nearly 96\% of Starlink's emissions are attributed to launch events, reflecting the intensive fuel combustion required for deployment. Conversely, smaller constellations, such as Lacuna, show a higher relative contribution from launcher production and electricity consumption, with launcher production accounting for 43.08\% of total emissions in some cases. Larger constellations, such as Starlink and Kuiper, align more closely with the average emission distribution, underscoring the influence of scale on emission profiles.

Emissions per rocket launch, displayed in Figure~\ref{fig:constellation_emissions}B, also demonstrate notable differences across constellations. Starlink records the highest emissions per launch at 5,658 t of CO$_2$, followed by HVNET at 2,500 t. In contrast, most other constellations exhibit emissions below 1,500 t per launch, reflecting variations in launch vehicle efficiency and payload capacity. For constellations like Kuiper, emissions contributions from ion propulsion systems (primarily arising from the production of krypton and xenon used in these systems) are significant, accounting for up to 13.3\% of total emissions. These findings highlight the critical role of launch vehicle selection and propulsion system design in shaping overall emissions efficiency.

When emissions are analyzed by impact category, both in Figure~\ref{fig:constellation_emissions}A and Figure~\ref{fig:constellation_emissions}B, considerable variability emerges. The launch event dominates total emissions for constellations such as Starlink, contributing up to 96.1\%, while for Kuiper, it constitutes a relatively smaller share at 25.5\%. Similarly, production-related emissions range from as little as 0.73\% of total emissions for Starlink to as high as 43.1\% for Lacuna, reflecting differences in deployment scales and satellite production efficiencies. Smaller constellations like Flock and Lacuna, with fewer launches, experience a disproportionate impact from production-related emissions, as these emissions constitute a larger fraction of their total footprint.

The influence of rocket reusability, shown in Figure~\ref{fig:constellation_emissions}D, also emerges as a significant factor in emissions reduction. Constellations utilizing reusable rockets, such as Starlink and New Glenn, show markedly lower emissions, averaging 735 t, with launcher production contributing only a small fraction of their total footprint (0.87\%). In contrast, constellations relying on non-reusable rockets exhibit higher emissions, averaging 7,487 tons, over 10 times that of reusable rockets. This is due to the repeated production of rocket stages and the larger contribution of launch events to their overall emissions profiles (67.4\%).

\subsection{Emissions Per User}

\begin{figure}[h]
    \centering
    \includegraphics[width=0.95\textwidth]{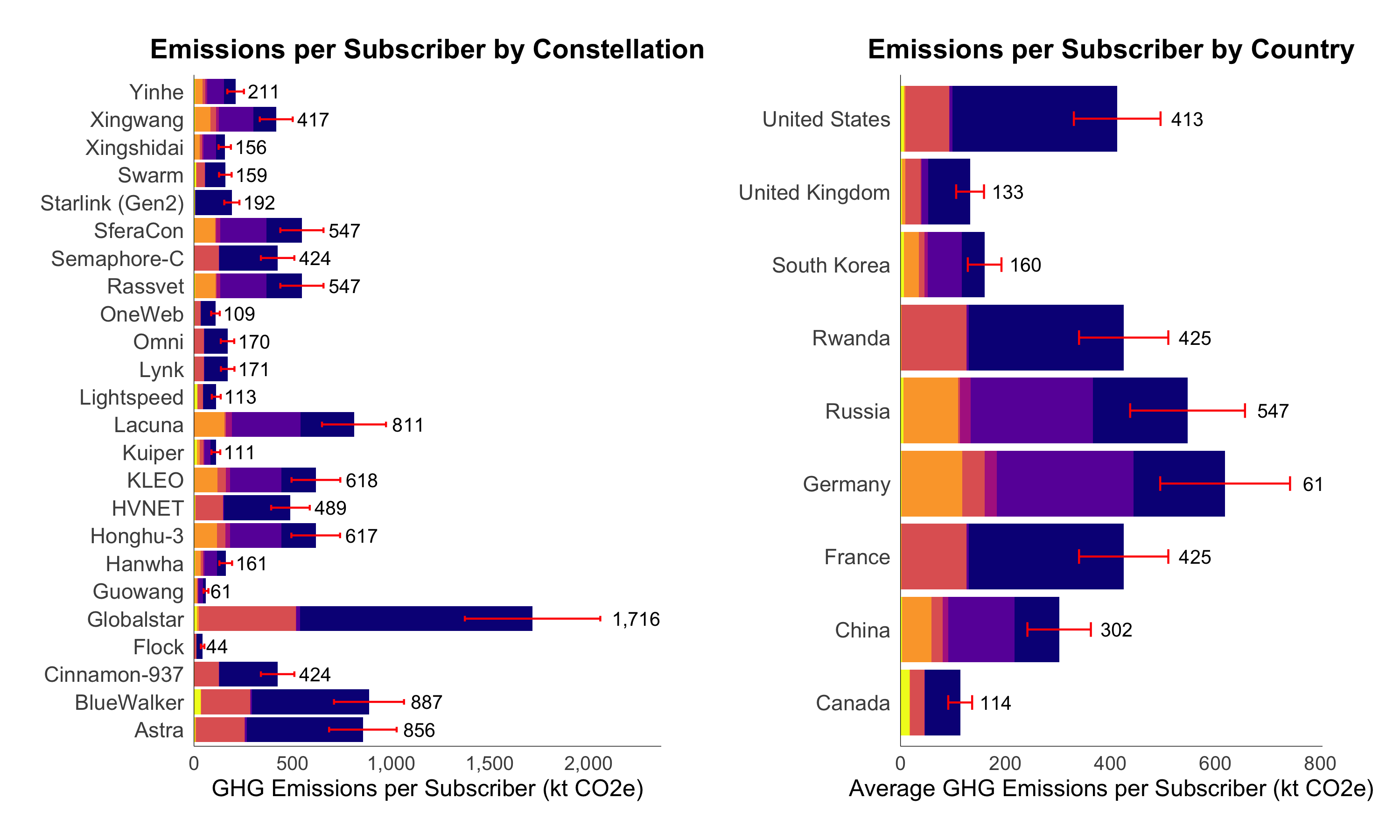}
    \caption{GHG emissions per subscriber, by satellite megaconstellation and by country.}
    \label{fig:subscriber_emissions}
\end{figure}

We also undertake a scenario analysis of emissions on a per-user basis for each constellation, shown in Figure~\ref{fig:subscriber_emissions}. Since the number of modeled subscribers varies greatly, ranging from 73,425 for Swarm and Flock to 165,119,609 for Cinnamon-937, it is important to adjust for this factor. Most constellations, except for Cinnamon-937, Semaphore-C, and Starlink, serve fewer than 7 million users. Adjusting emissions for user base size reveals key insights into the relative environmental efficiency of satellite constellations.

The average emissions per user is 417 ($\pm$ 74) kg, the equivalent of a flight from New York City, NY to Houston, TX \cite{kommenda_how_2019}. Figure~\ref{fig:subscriber_emissions}A shows that based on the methodology employed here, Globalstar stands out as the constellation with the highest emissions per subscriber, reaching a baseline of 1,716 ($\pm$ 394) kg of CO$_2$, a value more than 375\% above the average. This inefficiency arises from its relatively small user base and significant contributions from the launch event and propulsion systems. On the other hand, our scenario analysis indicates that the Flock constellation achieves the lowest emissions per subscriber at just 44 ($\pm$ 3) kg of CO$_2$, reflecting its highly efficient deployment and minimal emissions across all stages of its lifecycle. Notably, large constellations like Cinnamon-937 and Semaphore-C, despite their significant total emissions, are estimated to maintain relatively low per-subscriber emissions of 424 ($\pm$ 84) kg and 424 ($\pm$ 82) kg, respectively, due to their vast user bases. The distribution of emissions per user closely mirrors the total emissions breakdown by stage, with the launch event consistently dominating emissions for most constellations.

The emissions trends extend to a comparison of average per-subscriber emissions across countries in Figure~\ref{fig:subscriber_emissions}B. Based on  the analytical approach applied in this study, Germany records the highest emissions per subscriber, with an average of 618 ($\pm$ 119) kg of CO$_2$, 48.2\% above the average, largely driven by the inefficiency of the Globalstar constellation. Russia follows closely at 547 ($\pm$ 90) kg, reflecting a similar trend. By contrast, Canada and the United Kingdom are estimated to demonstrate significantly lower emissions per subscriber, at 114 ($\pm$ 21) kg and 133 ($\pm$ 32) kg, 72.6\% and 68.1\% lower than average, respectively, primarily due to their more efficient deployment methods.

\section{Discussion}\label{sec6}

Our comprehensive analysis of GHG emissions associated with satellite megaconstellations reveals critical insights into their environmental impact and highlights key areas for improvement. The results of this study provide valuable insights into the drivers of emissions at multiple scales — individual rockets, entire constellations, and on a per subscriber basis — effectively addressing all three research questions outlined in the Introduction.

The launch event and launcher transportation emerged as the dominant contributors, collectively accounting for 95.6\% of total emissions. This underscores the significant role of propellant combustion and transportation logistics in shaping the environmental footprint of satellite deployment. Addressing these emissions requires strategies such as optimizing rocket design, improving propellant efficiency, and adopting eco-friendly transportation methods.

Reusable rockets, such as SpaceX's Falcon 9 and Starship, represent a promising pathway for emissions reduction, with potential decreases in production-related emissions of up to 95.4\%. By avoiding the need to manufacture new rocket stages for each launch, reusable systems significantly lower emissions per event. However, challenges remain; for instance, Starship's large size results in the highest total emissions per launch, at 1.75 kt CO$_2$. Nonetheless, its emissions per kilogram of payload delivered are relatively low at 43.91 t CO$_2$, demonstrating its efficiency for large-scale deployments. Future advancements in lightweight materials, such as high-strength composites, and the integration of green propellants, such as methane and cryogenic hydrogen, are essential to further minimize the emissions footprint of reusable rockets.

The scenario variability in per-subscriber emissions across constellations introduces an important dimension of environmental efficiency. For example, our emissions analysis methodology estimates that given our assumptions, Globalstar produces 1,716 kg of CO$_2$ per subscriber—over 375\% higher than the average—due to its small subscriber base and high total emissions. In contrast, larger constellations like Cinnamon-937 and Semaphore-C achieve much lower per-subscriber emissions due to their extensive user bases and efficient scaling. For example, Cinnamon-937, which has the largest satellite network, achieves emissions of 424 kg CO$_2$ per subscriber. These findings highlight the importance of deploying constellations that balance satellite numbers and subscriber networks to reduce emissions per user.

Geographical differences in emissions profiles further illustrate the variability in sustainability practices. Constellations managed by Germany and Russia exhibit higher emissions per subscriber due to inefficient deployment strategies and smaller user bases. For instance, the emissions per subscriber for Russian constellations are approximately 547 kg CO$_2$, compared to a global average of 417 kg CO$_2$ per subscriber. In contrast, countries like Canada and the United Kingdom achieve much lower emissions per subscriber, associated with efficient constellations like Swarm and Flock. In fact, Flock demonstrates the lowest emissions per subscriber of any constellation, at just 44 kt CO$_2$. Additionally, the European Space Agency (ESA) demonstrates sustainable practices, including the use of eco-friendly transportation methods like the Canopée vessel, which reduces launcher transportation emissions by approximately 30\%. On the other hand, U.S.-produced rockets often lack comparable efficiencies in transportation logistics, leading to higher overall emissions.

The rapid growth of satellite megaconstellations poses significant long-term environmental risks. Without intervention, cumulative emissions from satellite launches could reach levels comparable to those of global subsonic aviation operations within the next decade. Additionally, the injection of particulates like alumina into the stratosphere could exacerbate radiative forcing and ozone depletion, further contributing to climate change. These risks underscore the urgent need for sustainable practices to ensure that the benefits of satellite constellations do not come at an unsustainable environmental cost.

Technological innovation offers promising solutions to these challenges. Optimizing flight trajectories and leveraging gravity assists could reduce propellant requirements by up to 60\%, while integrating eco-friendly propulsion systems, such as ion or electric propulsion, can minimize emissions during satellite station-keeping and orbital adjustments. Modular satellite architectures that extend operational lifespans and reduce the frequency of replacement launches represent another effective strategy for mitigating emissions.

The findings of this study emphasize the importance of adopting a holistic approach to sustainability in the satellite industry. Scaling subscriber networks, prioritizing rockets with lower emissions per payload, and incentivizing eco-friendly transportation methods are essential steps for reducing the environmental impact of satellite systems. International collaboration will be critical in establishing global standards for sustainable satellite deployment, ensuring alignment with global climate goals.

\section{Conclusion}\label{sec7}

This study provides a comprehensive LCA of GHG emissions associated with satellite megaconstellations.

This research contributes to the space sustainability field by providing critical insights into the primary drivers of emissions, such as propellant combustion and production processes. Additionally, it introduces the ORACLE project, which offers visualization tools and datasets to enable further research and collaboration. This open-source framework fosters transparency and innovation in studying space sustainability.

Our analysis reveals that launch events and launcher transportation dominate total emissions, collectively accounting for 95.6\% of the environmental impact. The average emissions per rocket launch are approximately 1.75 kt CO$_2$, equivalent to the annual carbon absorption of 80,000 trees. Reusable rockets demonstrate significantly lower production emissions—up to 95.4\% less than non-reusable alternatives—with production emissions accounting for only 1.05\% of total emissions compared to 26.2\% for non-reusable rockets. Per-subscriber emissions average 417 kg CO$_2$, comparable to a flight from New York to Houston, with large constellations achieving greater efficiency. For example, Cinnamon-937, which has the largest satellite network, has emissions of 61 kg CO$_2$ per subscriber, 85.3\% below average.

While comprehensive, this study has limitations. Assumptions about the number of satellites per rocket, material-specific emission factors, and reusability parameters may not fully capture the diversity of satellite constellations and launch systems. Future research should refine these parameters, incorporate real-world operational data, and explore emerging technologies such as kinetic launch systems and alternative propulsion methods. Further investigation into the cumulative effects of launches on atmospheric chemistry and orbital debris is also necessary to ensure the long-term sustainability of satellite constellations.

A key contribution of this study is its inclusion of a broader range of international launch vehicles in environmental analyses. This approach provides a more comprehensive understanding of the global impact of satellite constellations, addressing a significant gap in previous research. By expanding the scope of analysis beyond U.S. and European launch vehicles to include rockets developed by other nations, this study offers a more accurate representation of the emissions profile associated with global satellite deployment.

Balancing the transformative benefits of satellite megaconstellations—such as global connectivity and advancements in communication technologies—with their environmental impacts is critical. Strategies like optimizing flight trajectories, leveraging gravity assists, and adopting green propulsion systems can significantly reduce emissions. Modular satellite designs that extend operational lifespans and reduce replacement frequency also offer a viable pathway to sustainability.

International collaboration and policy development will further play a vital role in promoting sustainable growth within the space industry. Establishing global standards for satellite deployment, encouraging reusable launch systems, and incentivizing eco-friendly practices will ensure that the expansion of satellite constellations aligns with global sustainability goals.

In conclusion, this study highlights the need to integrate environmental sustainability into the design, development, and deployment of satellite megaconstellations. By addressing key areas such as reusability, emissions efficiency, and innovative technologies, the space industry can balance technological progress with environmental stewardship, paving the way for a sustainable future in satellite deployment and space exploration.

\bibliography{references}

\end{document}